\documentclass[aps,prl,twocolumn,superscriptaddress]{revtex4-1}

\usepackage{graphicx}
\usepackage{bm}
\usepackage{ulem}
\usepackage{color}

\begin{document}

\title{Nonclassical optical properties of mesoscopic gold}

\author{Swen~Gro{\ss}mann}
\author{Daniel~Friedrich}
\affiliation{Nano-Optics and Biophotonics Group, Department of Experimental Physics 5, Wilhelm-Conrad-R\"ontgen-Center for Complex Material Systems (RCCM), University of W\"urzburg, Am Hubland, 97074 W\"urzburg, Germany}
\author{Michael~Karolak}
\affiliation{Institute for Theoretical Physics and Astrophysics, University of W\"urzburg, Am Hubland, 97074 W\"urzburg, Germany}
\author{Ren\'e~Kullock}
\author{Enno~Krauss}
\author{Monika~Emmerling}
\affiliation{Nano-Optics and Biophotonics Group, Department of Experimental Physics 5, Wilhelm-Conrad-R\"ontgen-Center for Complex Material Systems (RCCM), University of W\"urzburg, Am Hubland, 97074 W\"urzburg, Germany}
\author{Giorgio~Sangiovanni}

\affiliation{Institute for Theoretical Physics and Astrophysics, University of W\"urzburg, Am Hubland, 97074 W\"urzburg, Germany}
\author{Bert~Hecht}
\email[]{hecht@physik.uni-wuerzburg.de}

\affiliation{Nano-Optics and Biophotonics Group, Department of Experimental Physics 5, Wilhelm-Conrad-R\"ontgen-Center for Complex Material Systems (RCCM), University of W\"urzburg, Am Hubland, 97074 W\"urzburg, Germany}

\begin{abstract}

Gold nanostructures have important applications in nanoelectronics, nano-optics as well as in precision metrology due to their intriguing opto-electronic properties. These properties are governed by the bulk band structure but to some extend are tunable via geometrical resonances. Here we show that the band structure of gold itself exhibits significant size-dependent changes already for mesoscopic critical dimensions below 30~nm. To suppress the effects of geometrical resonances and grain boundaries, we prepared atomically flat ultrathin films of various thicknesses by utilizing large chemically grown single-crystalline gold platelets. We experimentally probe thickness-dependent changes of the band structure by means of two-photon photoluminescence and observe a surprising 100-fold increase of the nonlinear signal when the gold film thickness is reduced below 30~nm allowing us to optically resolve single-unit-cell steps. The effect is well explained by density functional calculations of the thickness-dependent 2D band structure of gold.
\end{abstract}
%

\maketitle
The emergence of 2D semiconductors with novel, unexpected properties has sparked a renewed interest in thickness-dependent properties of ultrathin films. For gold, such studies have started two decades ago by investigating few-atomic-layer-thick films. In such structures the electrons are found to be strongly confined. This results in discrete quantum-well states that dominate the dielectric  \cite{dryzek_quantum_1987,vyalykh_quantum-well_2002,romanishin_quantum-well_1999} and electronic properties, and lead e.g. to quantized electrical conductance \cite{rodrigues_signature_2000,pascual_properties_1995}. However, due to the fcc lattice the required gold mono- or multilayers could not be easily prepared, e.g. by exfoliation. Therefore, experimental efforts towards exploiting such effects gradually phased out. In the meantime, gold nanostructures became a topic of intense research and their optoelectronic properties have enabled nonlinear optical \cite{cai_electrically_2011,aouani_third-harmonic-upconversion_2014,celebrano_mode_2015} and hot-electron-induced \cite{knight_photodetection_2011,fang_graphene-antenna_2012,christopher_visible-light-enhanced_2011,linic_plasmonic-metal_2011} functionality, strong light-matter coupling at ambient conditions \cite{chikkaraddy_single-molecule_2016,gross_near-field_2018} as well as new sensing technology \cite{nylander_gas_1982,wu_refractive_2009}. For gold structures below 10~nm the influence of the finite size on the material is well known \cite{scholl_quantum_2012}. In the mesoscopic regime - with the exception of electron-surface-scattering to account for line broadening \cite{berciaud_observation_2005,carmina_monreal_competition_2013} - commonly the bulk properties are used to describe and model the geometry-dependent opto-electronic properties.\newline
\begin{figure}[bp]
	\centering
 	\includegraphics[width=0.47\textwidth]{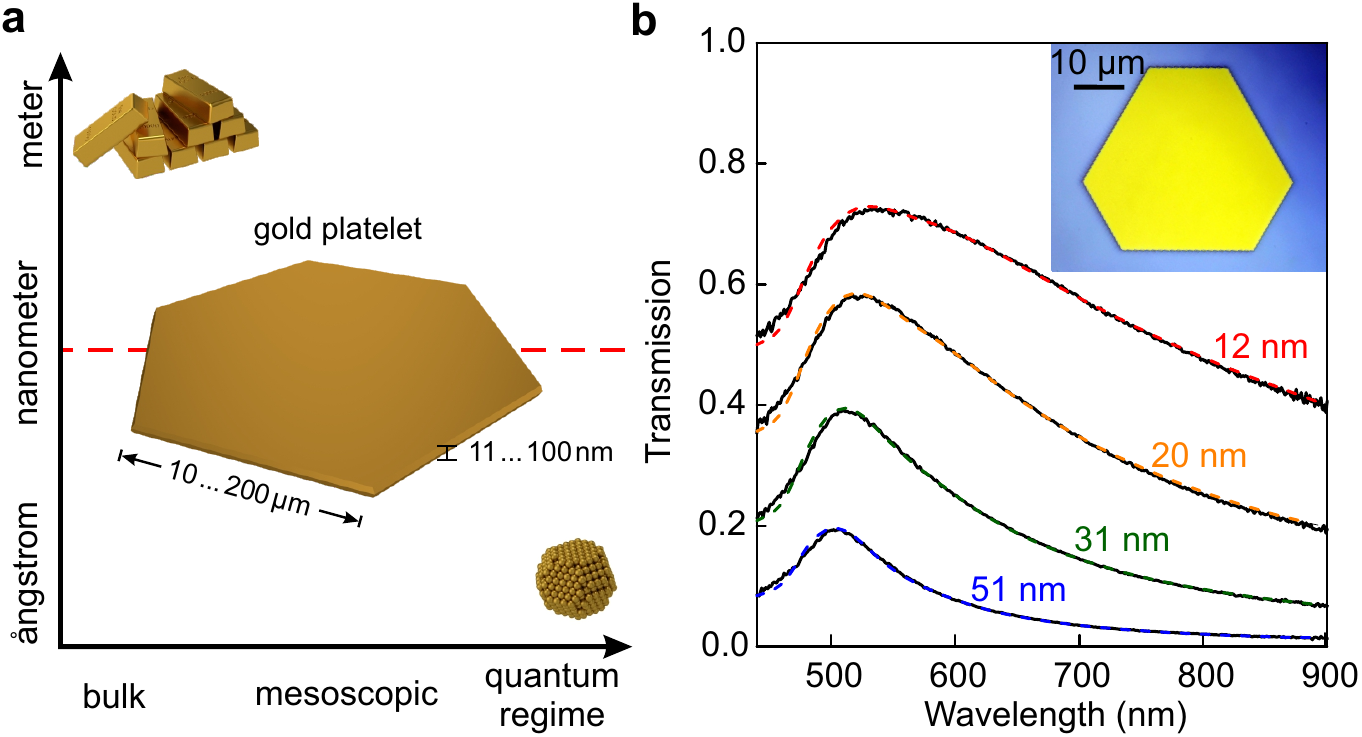}
 	\caption{\label{fig:1} (a) The size of gold structures in comparison to the transition between the full classical (bulk) and full quantum regime, including the electron mean free path (red dashed line). (b) Transmission spectra (solid black lines) of four gold microplatelets fitted for thicknesses of 12~nm (dashed red), 20~nm (dashed orange), 31~nm (dashed green), and 51~nm (dashed blue). The inset shows a true-color optical reflection microscopy image of a typical gold microplatelet.}
\end{figure}
\begin{figure*}[hbtp]
	\centering
 	\includegraphics[width=\textwidth]{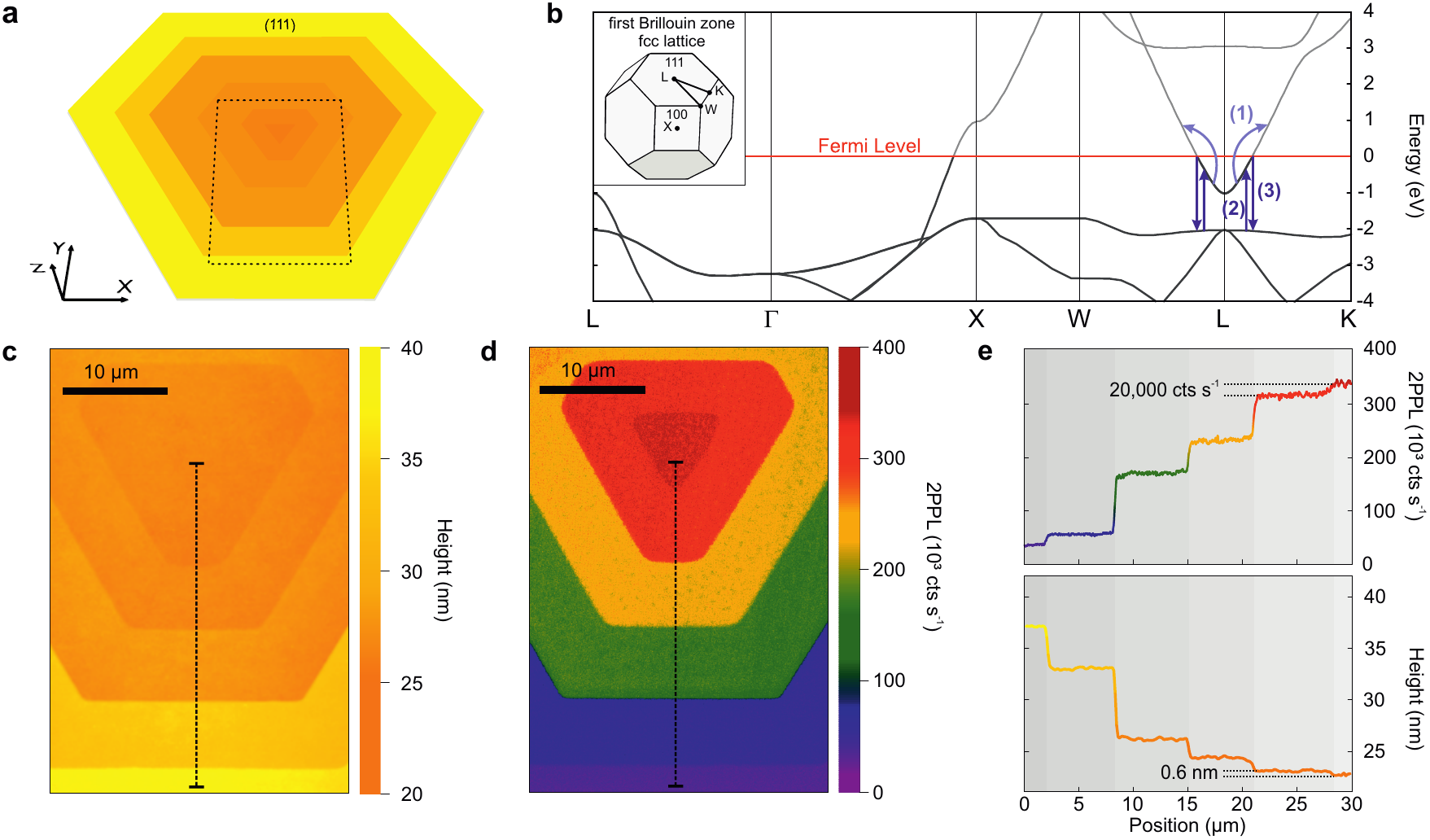}
 	\caption{\label{fig:2} (a) Illustration of a terraced gold microplatelet including the investigated area. The upper surface corresponds to the (111) crystal plane. (b) Two-photon photoluminescence process, illustrated within the band structure of bulk gold (obtained from DFT calculations). The excitation wavelength is 816~nm. The inset shows the first Brillouin zone of the face-centered cubic lattice. (c) Atomic force micrograph of a terraced microplatelet (area highlighted in (a)). (d) Two-photon photoluminescence scan image of the same platelet as in (c). (e) The height profile marked in (c) and the intensity profile from (d).}
\end{figure*}
Here we show that the band structure of gold is significantly altered with respect to bulk already for gold film thicknesses of less than 30~nm which for example is a typical size range of plasmonic nanostructures. Changes manifest themselves in the appearance of new bands in the conduction band regime and corresponding direct dipolar interband transition within the conduction band regime. We demonstrate that two-photon photoluminescence (2PPL) provides an elegant way to probe such band structure alterations with very high sensitivity. We show that 2PPL can resolve thickness variations in gold films on the length scale of single atoms. \newline
To exclude any influence of geometrical resonances or grain boundaries we prepare atomically flat films of various thicknesses by utilizing single-crystalline gold platelets. They are chemically synthesized on glass substrates by immersing microscope cover slips into a growth solution: $\mathrm{HAuCl_4}$ + $\mathrm{H_2O}$ + ethylene glycol (Supplemental Material \cite{supplemental_material} for details). By means of heat-supported self-assembly, triangular as well as hexagonal platelets emerge \cite{krauss_controlled_2018}, exhibiting thicknesses between 11 and 100~nm and lateral dimensions of up to 200~$\mu$m (cf.~Fig.~\ref{fig:1}). Due to the enormous lateral dimensions, the microplatelets do not support localized plasmon resonances in the investigated wavelength range and thus can be considered as quasi-infinite gold films, which are single-crystalline except of a few stacking faults, i.e. twin planes. Not all of the platelets are completely flat, but some interestingly show a terrace-like structure as depicted in Fig.~\ref{fig:2}a. The observed terrace step heights vary between a few hundred picometers, corresponding to a few atomic layers, to tens of nanometers. Consequently, very high-precision thickness-dependent measurements can be conducted by using a piezo stage in order to scan the sample. Importantly, throughout the single-crystalline gold platelet the influence of electronic states emerging at interfaces due to the broken translational symmetry \cite{reinert_spinorbit_2003} is kept constant.\newline
First we characterize the linear optical properties of gold microplatelets by very carefully recording transmission spectra for various thicknesses ranging from 51 to 12~nm (see Fig.~\ref{fig:1}b and Supplemental Material \cite{supplemental_material} for more details). Surprisingly, the measured spectra show no visible deviations from calculated transmission spectra obtained with the bulk dielectric constant of gold \cite{olmon_optical_2012} even for the smallest thickness. While it was speculated, that there may be some deviations from the bulk electronic band structure \cite{etchegoin_analytic_2006, beversluis_continuum_2003, yakubovsky_optical_2017}, it seems that linear white-light transmission spectroscopy is not sensitive enough to reveal such thickness-dependent effects since it mainly probes d-band to conduction band interband transitions and the free electron density at the Fermi edge.\newline
\begin{figure*}[!hbt]
	\centering
 	\includegraphics[width=\textwidth]{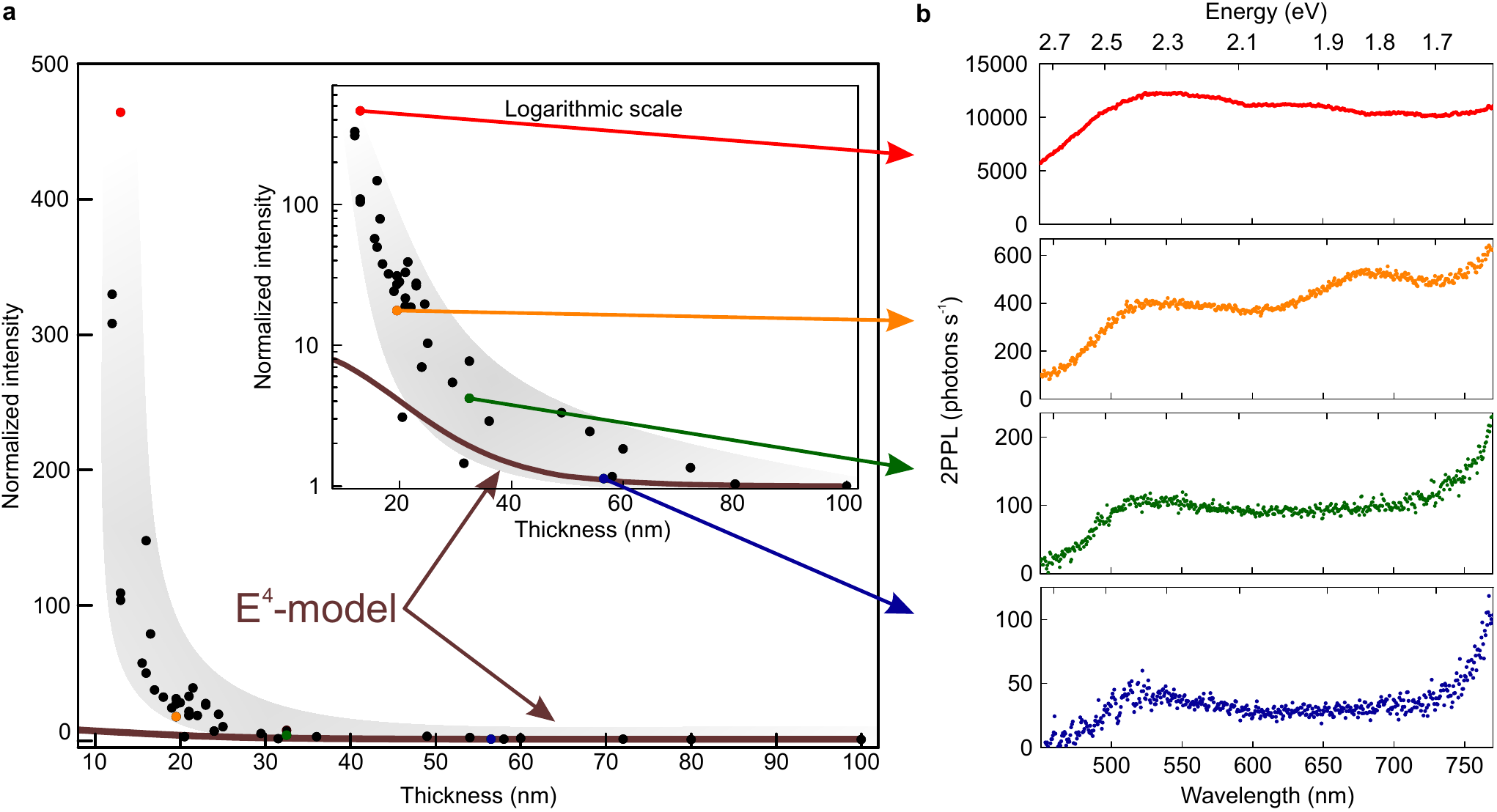}
 	\caption{\label{fig:3}(a) Thickness dependence of the 2PPL signal. The inset shows the same plot using logarithmic scale. The brown line shows the expected intensity according to formula (1). (b) 2PPL spectra of gold platelets for a thickness of 13~nm (red), 19~nm (yellow), 32~nm (green), and 56~nm (blue).}
\end{figure*}
A much more versatile and subtle probe of band structure variations as a function of the gold layer thickness, which is also well suited to resolve thickness variations of single atomic layers in 2D luminescent materials \cite{liu_nonlinear_2010} with diffraction limited lateral resolution \cite{viarbitskaya_tailoring_2013}, is nonlinear 2PPL spectroscopy. The extreme sensitivity of 2PPL with respect to band structure variations away from the Fermi edge roots in the fact that it is a cascaded second-order nonlinear process consisting of three steps (Fig.~\ref{fig:2}b) \cite{beversluis_continuum_2003, imura_near-field_2005, zijlstra_single_2011, biagioni_dynamics_2012}. First, in the conduction band, hot electrons are excited via intraband transitions thereby creating vacancies below the Fermi edge. In a second step, these vacancies are filled by d-band electrons. Finally, the thus created d-band holes are filled via interband transitions. The resulting excess energy converts into far-field photons either directly or via surface-plasmon polaritons, which may lead to a characteristic spectral shaping of 2PPL \cite{beversluis_continuum_2003}. In absence of a distinct plasmonic resonance, the emitted photons exhibit the energy of the interband transitions involved in the last step \cite{imura_near-field_2005}. It is important to note that due to the fact that the first step involves intraband transitions, which require a source of momentum to enable the electrons changing their energy, 2PPL in noble metals is expected to only occur in nanostructured materials or materials with significant roughness because of the resulting strong light confinement \cite{huang_atomically_2010} and to be absent in flat thin films \cite{melentiev_giant_2015}. \newline
To investigate the 2PPL we used the terrace-shaped platelets. The thickness of this type of gold platelet is always increasing from the center towards the rim as illustrated in the atomic force micrograph (Fig.~\ref{fig:2}c). Importantly, the smallest obtained gold thickness of 11~nm, corresponding to about 46~atomic~layers, is in the range of thicknesses for which quantum confinement and surface effects are expected to become significant. \newline
By comparing the AFM image and the corresponding 2PPL map shown in Fig.~\ref{fig:2}d, three main features become apparent: First, surprisingly, significant 2PPL signal is observed despite the flatness of the terraces which strongly increases with decreasing thickness of the gold. Second, regions of the gold platelet with the same thickness, i.e. the terraces, exhibit a very homogenous signal, providing additional evidence for the extreme small surface roughness and flatness of the terraces \cite{beversluis_continuum_2003} (cf. Supplemental Material~\cite{supplemental_material}). Finally, we do not observe an increase of the signal on the rim of the terraces, concluding there is no enhancement of the 2PPL due to the edges of the terraces. Fig.~\ref{fig:2}e highlights the smallest discrete step that we found on a gold microplatelet, corresponding to the height of one unit cell. This height of only 0.6~nm, which also indicates the small surface roughness, can clearly be resolved in the 2PPL measurement. Thus, 2PPL turns out to be a powerful tool to map thickness variations on the atomic scale not only for 2D materials like graphene but also for noble metals, i.e. gold. \newline
In order to further investigate the origin of this behavior we study the spectrally integrated 2PPL intensity as a function of the plate thickness (Fig.~\ref{fig:3}a). Here we observe a completely unexpected increase of the 2PPL intensity by two orders of magnitude when decreasing the gold thickness from 100~nm to 11~nm. To explore the mechanism behind this increase, 2PPL was spectrally dispersed (Fig.~\ref{fig:3}b). Above 750~nm all spectra show a wavelength-dependent increase of the signal. We assume this to be the residual of the excitation laser (center wavelength of 816~nm and pulse duration of 120~fs), which contributes a thickness-independent amplitude of approximately 100~photons/s to each spectrum. All spectra exhibit a maximum at 520~nm which corresponds to the interband transitions near the L-symmetry point in the first Brillouin zone of gold (Fig.~\ref{fig:2}b) \cite{imura_near-field_2005}. However, when changing the thickness, apart from the increased 2PPL amplitude, the spectral shape gets modified. This becomes noticeable at plate thicknesses below 30~nm and continues for decreasing thickness (Fig.~\ref{fig:3}b). It should be mentioned here, that the spectral shape is independent of the excitation power, unless the damage threshold is reached (cf.~Supplemental Material~\cite{supplemental_material}). The same thickness dependent behavior of the 2PPL is found for gold platelets without terraces. \newline
One may speculate that the observed effects are solely caused by the excitation of surface plasmon polaritons. These near-field modes exhibit large fields inside the metal and therefore represent an important decay channel for electronic excitations in gold. To explore the possible influence of surface plasmons, we recorded the 2PPL radiation pattern (cf. Supplemental Material \cite{supplemental_material}). A distinct ring-like intensity distribution beyond the critical angle of total internal reflection at the glass-air interface is observed which clearly signifies a contribution of the surface-plasmon-polariton channel to the radiative decay. In a control experiment we therefore suppressed the far-field emission of gold-air surface plasmons by introducing an index matching material on top of the gold platelets thereby creating a symmetric environment. However, even in absence of surface plasmon-mediated decay channels, the gold platelet 2PPL still exhibits the same spectral distribution and a similar thickness dependence of the integrated 2PPL signal. We therefore exclude the enhanced excitation of surface plasmon polaritons for decreasing plate thickness as the origin of the increasing 2PPL signal.\newline
Next, we investigate the influence of electromagnetic effects, such as the thickness-dependent field penetration and localization in the gold slab. We therefore calculate the expected relative change of the 2PPL intensity based on the intensity of excitation light inside the metal layer. The established method to approximate 2PPL assumes the intensity to be proportional to the fourth power of the local electric field inside the gold film generated by the laser beam at the excitation frequency \cite{viarbitskaya_tailoring_2013, boyd_photoinduced_1986, huang_mode_2010, teulle_scanning_2012, ghenuche_spectroscopic_2008}:
\begin{equation}
I_{TPPL}(\omega_0) = \eta^2(\omega_0)\int_V \big| \mathbf{E}(\omega_0,\textbf{r}) \big|^4 d\textbf{r}. \label{eq:1}
\end{equation}
Here $\mathbf{E}$ denotes the electric field, $\omega_0$ describes the excitation frequency, $V$ is the volume occupied by gold, and $\eta$ is a frequency dependent material parameter. Note that this widely used and quite successful method does not correctly account for momentum conservation, i.e. the need of considering intensity gradients as explained above. However, the use of a modified formula which includes field gradients yields similar results (cf. Supplemental Material~\cite{supplemental_material}) justifying the simplified approach of Eq. (\ref{eq:1}).\newline
To compute the field in the gold platelets they are illuminated with a focused Gaussian beam in finite-difference time-domain simulations (cf. Supplemental Material \cite{supplemental_material}). By explicitly calculating the 2PPL intensity (brown line, Fig.~\ref{fig:3}a) the simulations predict an increase by a factor of eight in the 2PPL signal, when decreasing the gold thickness from 100~nm to 10~nm. In contrast, the experimental data show a more than 100-fold 2PPL-enhancement for the thinnest platelets. The strong increase of 2PPL towards ultrathin gold platelets as well as the concomitant pronounced spectral modification clearly indicate a thickness-dependent change in the optical properties of gold.\newline
To specify the changes occurring for ultrathin gold films, we performed DFT calculations of the electronic band structure and the resulting optical properties of gold (cf. Supplemental Material \cite{supplemental_material}). The calculations were performed using the all-electron full-potential APW+lo method and the PBE-GGA functional \cite{schwarz_solid_2003}. The crystal structures, for which computations were performed, are illustrated in Fig.~\ref{fig:4}a. They consist of perfect face-centered cubic slabs except of two (or more, see Supplemental Material \cite{supplemental_material}) stacking faults at which the stacking order is mirrored. These twin planes are oriented parallel to the (111) crystal plane as expected for single-crystalline gold microplatelets \cite{hoffmann_new_2016}. The existence of twin planes in the crystal is mainly causing a shift of the band energies as well as changes in the dipolar matrix elements (cf. Supplemental Material \cite{supplemental_material}). Based on the band structure for a 10~nm gold film, which is displayed in Fig.~\ref{fig:4}b, we calculated the interband contribution to the imaginary part of the dielectric tensor by only considering the contribution of dipolar interband transitions \cite{ambrosch-draxl_linear_2006}:
\begin{widetext}
\begin{equation}
\varepsilon^{(inter)}_{ij}(\omega) = \frac{\hbar^2 e^2}{\pi m^2 \omega^2}\sum_{n,n'}\int d \boldsymbol{k} ~ p_{i;n',n,\boldsymbol{k}} ~ p_{j;n',n,\boldsymbol{k}} ~ \lbrack f(\epsilon_{n,\boldsymbol{k}} - \epsilon_{n',\boldsymbol{k}}) \rbrack ~ \delta(\epsilon_{n,\boldsymbol{k}} - \epsilon_{n',\boldsymbol{k}} - \omega).
\end{equation}
\end{widetext}
Here, $\epsilon_{n,\boldsymbol{k}}$ are the band energies and $p_{i;n',n,\boldsymbol{k}}$ are the matrix elements of the momentum operator, what remains is the joint density of states. We thus obtain a quantity which is proportional to the absorption of light in gold via direct interband excitation of electrons. 
\begin{figure*}[htp]
	\centering
 	\includegraphics[width=\textwidth]{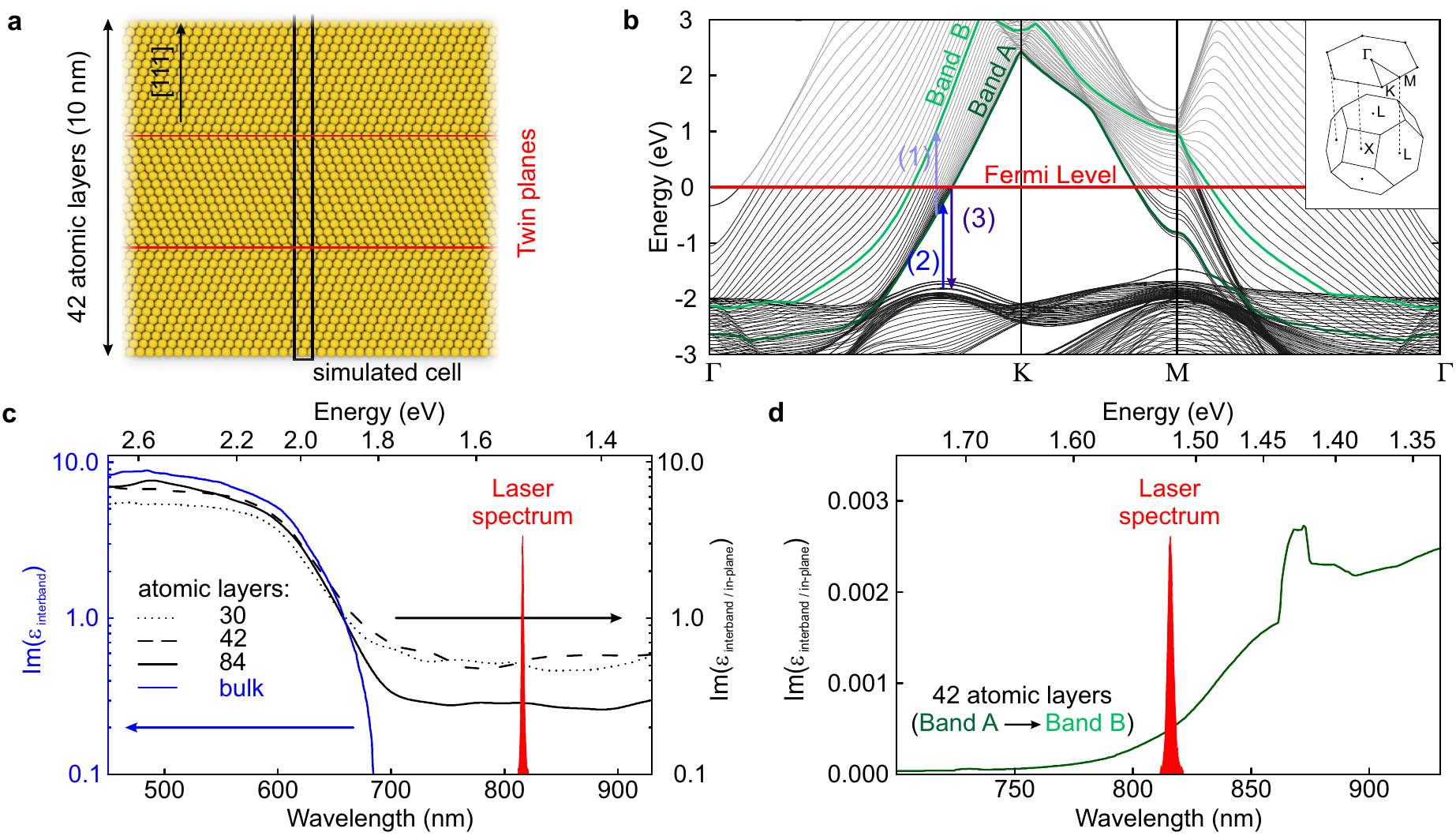}
 	\caption{\label{fig:4} (a) The calculated 10~nm (42 atomic layers) gold slab terminated by vacuum at the top and bottom, the two twin planes and a part of the simulated cell. (b) Energy bands of a 10~nm thick gold slab. The arrows illustrate one possible realization of a modified nonlinear photoluminescence process. The inset shows the face-centered cubic Brillouin zone as well as the Brillouin zone of a single atomic layer of gold. (c) The imaginary part of gold resulting from dipole interband transitions for bulk (solid blue) as well as the in-plane component for 84 (solid black), 42 (dashed black), and 30 (dotted black) atomic layers. The red peak indicates the excitation laser, used in the experiments. (d) In-plane component of the imaginary part of the dielectric function resulting from interband transitions solely between Band A and Band B in the 42-layers gold slab. 
}
\end{figure*}
As shown in Fig.~\ref{fig:4}c the interband transitions exhibit a large contribution for wavelengths shorter than 600~nm for bulk as well as for the computed slab geometries. For bulk gold the contribution of dipole allowed interband transitions vanishes above 700~nm. In contrast for thin gold slabs we observe a strong contribution of dipole-allowed electronic interband transitions above 700~nm which increases as the gold thickness decreases. The twin planes enhance these changes to the dielectric tensor, but their presence is not a necessary condition for the observed phenomenon, hinting at quantum confinement as its origin \cite{ehlen_evolution_2016}. A similar increase of the contribution of allowed electronic interband dipole transitions for decreasing thickness is also observed for a perfect (111) slab (cf. Supplemental Material \cite{supplemental_material}). These observations are consistent with the perception that 2PPL is a very inefficient process in bulk gold due to the absence of dipole-allowed interband transitions for the first step of the 2PPL excitation cascade. In marked contrast, ultrathin single-crystal thin films do possess ample possibilities to excite interband transitions within the conduction bands leading to the observed tremendous enhancement of the 2PPL process. \newline
To illustrate the occurrence of interband transitions at the laser wavelength of 816~nm and thus the modification of the 2PPL process for thin gold films, we exemplarily chose two bands, Band A and Band B, from the whole range of bands that contribute to interband transitions for a film thickness of 10~nm. Fig.~\ref{fig:4}b illustrates a possible sequence of transitions resulting in 2PPL emission at 680~nm. The imaginary part of the contribution to the dielectric function that is solely caused by the dipolar transitions between Band A and Band B is plotted in Fig.~\ref{fig:4}d. A clear non-zero contribution at the laser wavelength and an increase towards the infrared region can be observed, which can explain the enhanced 2PPL efficiency due to dipolar excitations in the first step of the process. Solely taking light confinement into account provides less than one tenth of the observed 2PPL enhancement and therefore cannot explain the effect (cf. Supplemental Material Fig.~S14 \cite{supplemental_material}). We assume that the spectral modification for thinner platelets originates from the large amount of possible transitions involved in the 2PPL process for thin gold films. Our results not only explain the huge enhancement of the nonlinear photoluminescence, but also previous findings of linear photoluminescence of thin gold films in the near infrared \cite{beversluis_continuum_2003}.\newline
In conclusion, for plate thicknesses smaller than 30~nm we observe a huge increase in the 2PPL signal as well as a concomitant effect on the dielectric function. Analyses of both the electromagnetic fields and the electronic structure of thin gold slabs show that this signal increase is caused by an arising two-dimensional character of gold for thicknesses below 30~nm accompanied by characteristic modifications of the band structure caused by the finite film thickness and the thus broken crystal periodicity. We show that 2PPL of gold can even be used to measure variation in the thickness down to a single unit cell. Our results concerning the dielectric properties of thin gold slabs and their modified optical response open up new possibilities and perspectives for the design of light harvesting and nonlinear frequency conversion devices. The atomically flat single-crystalline gold microplatelets in combination with the use of nonlinear photoluminescence to determine the thickness may also be useful in high-precision metrology, for example in Casimir force experiments \cite{rodriguez_casimir_2011}. 
%
%
\begin{acknowledgments}
We thank Peter Blaha for his help with WIEN2k. G.S. acknowledges useful discussions with Gianni Profeta. The authors gratefully acknowledge the Gauss Centre for Supercomputing e.V. (www.gauss-centre.eu) for funding this project by providing computing time on the GCS Supercomputer SuperMUC at Leibniz Supercomputing Centre (LRZ, www.lrz.de). This work was supported by the DFG through FOR 1162 (M.K.) and SFB 1170 ‘ToCoTronics’ (G.S.).
\end{acknowledgments}
\end{document}